\documentclass[twocolumn,draft,aps,preprintnumbers,amsmath,amssymb,nofootinbib,]{revtex4}

\usepackage{graphicx}
\usepackage{dcolumn}
\usepackage{bm}
\usepackage{amssymb,amsmath,amsfonts,amsbsy,epsfig}
\usepackage{color}

\def\lsim{\mathrel{\rlap{\lower4pt\hbox{\hskip1pt$\sim$}}
    \raise1pt\hbox{$<$}}}         
\def\gsim{\mathrel{\rlap{\lower4pt\hbox{\hskip1pt$\sim$}}
    \raise1pt\hbox{$>$}}}
\def\fun#1#2{\lower3.6pt\vbox{\baselineskip0pt\lineskip.9pt
  \ialign{$\mathsurround=0pt#1\hfil##\hfil$\crcr#2\crcr\sim\crcr}}}

\input epsf

\begin{document}

\preprint{\tt IPMU10-0190, \,\,\, \\ \today}

\title{Hyperbolic Inflation}


\author{Yoonbai Kim${}^{1}$} 
\email{yoonbai@skku.edu}
\author{Seong Chan Park${}^{2}$}
\email{seongchan.park@ipmu.jp}
\affiliation{${}^{1}$Department of Physics, BK21 Physics Research Division,
and Institute of Basic Science,\\
Sungkyunkwan University, Suwon 440-746, Korea\\
${}^{2}$Institute for the Physics and Mathematics of the Universe, 
The University of Tokyo, Chiba $277-8568$, Japan}

\vspace{1.0cm}
\begin{abstract}
We propose a natural scenario for the cosmological inflation with the non-minimal coupling term invoking compact hyperbolic extra dimensions. 
Thanks to the unique mathematical properties of compact hyperbolic space,  the large volume of extra dimensions, which provides a natural understanding of the proper size of couplings,  does not necessarily accompany with  the low Kaluza-Klein scale so that the model allows a single field inflation with a scale around $10^{13}$ GeV. The model fulfills all the observed data and predicts a sizable gravitational perturbation, $r\simeq 3\times 10^{-3}$. 
\end{abstract}

\maketitle

\section{Introduction}


A compelling idea is the presence of the inflationary era in the early universe \cite{inflation_original}. Inflation solves several cosmological problems and its generic predictions 
are consistent with various observations \cite{text}. Recently inflation models based on the non-minimal (NM) coupling term, the direct coupling term between a scalar field and the Ricci scalar, $\phi^2 R$,  have drawn a sizable attention among physicists and cosmologists \cite{history} as the standard model Higgs field is claimed to play the role of inflation without invoking additional field contents \cite{Bezrukov:2007ep}. 

Indeed the underlying mechanism of the NM inflation is shown that the slow-roll potential for the proper inflationary era is quite generically obtainable provided that the ratio between the potential in Jordan frame and the NM coupling term, $V(\phi)/K^2(\phi)$, is asymptotically constant \cite{Park:2008hz} where $V (K)$ is the potential (NM coupling term) in Jordan frame, respectively. One should note that the appearance of the NM coupling term is allowed in supergravity theories taking generic K\"ahler potential into account \cite{Ferrara:2010yw}.  Also the quadratic NM term $\sim \phi^2 R$ is of the same order of the leading Einstein-Hilbert action so that one should consider the NM term  in the effective field theory point of view \cite{Georgi:1994qn}. 

Even though the class of models with the NM term provides successful inflation, 
a fine-tuning problem, however, appears in fitting the observed temperature fluctuation in Cosmic Microwave Background Radiation (CMBR), $\Delta T/T\sim 10^{-5}$ quite generically \cite{Park:2008hz}. For instance, when $V=\lambda \phi^4$ and $K=\xi \phi^2$, the ratio between the coupling constants $\lambda/\xi^2$ is required to be extremely small as the value is determined by the precisely measured value of  primordial density perturbation by WMAP \cite{wmap7},
$\delta_H^2 =\frac{1}{75\pi M_4^6}\frac{U_{{\rm E}}^3}{U_{{\rm E}}'}\approx (1.91 \times 10^{-5})^2$:
\begin{eqnarray}
\frac{\lambda}{\xi^2}\approx 
4.4 \times 10^{-10}. 
\label{eq:effective}
\end{eqnarray}
Since the smallness is not a consequence of a symmetry of the model, we would regard the appearance as a fine-tuning problem in the 't Hooft sense \cite{'tHooft:1979bh}.  
The main purpose of the current paper is to address this fine-tuning problem by extending the model to the higher dimensions with compact hyperbolic space (CHS) as extra dimensions. 

For a quite long time, mathematicians \cite{Thur} have noticed that CHS has a unique property, dubbed, {\it rigidity} after G. Mostow \cite{Mosto} and the mass gap on CHS is conjectured to be greater than $1/2$ in the unit of the curvature scale after A. Selberg \cite{Selb, Luo, Cornish:1997an}. Relatively recently in physics there have been works  focused on phenomenoloy with a rather low KK scale ($\sim$ TeV)~\cite{Kaloper:2000jb, Demir:2001ap, Melbeus:2008hk, Orlando:2010kx} and a compactification scheme in string theory~\cite{Kehagias:2000dga, Chen:2003dca, Douglas:2010rt}. Also there have been works considering CHS in the context of cosmology \cite{Kaloper:2000jb}, in particular, inflation by embedding the chaotic inflation model  into higher dimensions with CHS without considering NM term \cite{Greene:2010ch}. The attempt in Ref.~\cite{Greene:2010ch}, however, still requires further fine-tuning as the simple model with $\lambda \phi^4$ potential does not fit the recent WMAP data in $(n_{\rm s}-r)$ plane within $2\sigma$ confidence level \footnote{Indeed a phenomenologically modified potential $V=\lambda e^{2\alpha \phi^2/v^2}(\phi^2 - v^2)^2$ is tried in \cite{Greene:2010ch} to reduce the deviation.}.  In this regard, one may consider the model in current paper as a successful refinement of the model. 

The rest of this paper is organized as follows.
In section II we review the unique mathematical properties of CHS which allows us to consider a big volume without a low mass gap.  In section III we recapitulate the inflationary models with NM coupling in $(1+3)$ dimensions and show the fine-tuning problem in detail. 
In section IV, finally, we extend the model in higher dimensions with CHS and show how
the fine-tuning problem is solved. We conclude in section V with some comments on consistency of the proposed model.

\section{Compact hyperbolic space}

Hyperbolic space in $d$ dimensions, ${\cal H}^{d}$ $(d\ge 2)$, is visualized as 
a hyperboloid embedded in $1+d$ dimensions, keeping SO$(d,1)$ symmetry manifest,
\begin{align}
-x_{0}^{2}+x_{1}^{2}+x_{2}^{2}+\cdots +x_{d}^{2}=-\ell^{2},
\end{align}
where $\ell$ represents the length scale from the negative curvature $R$
\begin{align}
R=-\frac{d(d-1)}{\ell^{2}}.
\label{cu1}
\end{align}
Its source can be a negative cosmological constant $\Lambda$ for $d\ge 3$, 
\begin{align}
\Lambda=-\frac{(d-1)(d-2)}{2\ell^{2}}.
\label{cc1}
\end{align}
In terms of a radial variable $\chi~(|\chi|\ge \ell)$ and $d-1$ angular variables 
$\theta_{a}$,
the induced metric can be expressed by
\begin{align}
ds_{d}=&\ell^{2}(d\chi^{2}+\sinh^{2}\chi d\Omega_{d-1}^{2}),\\
d\Omega_{d-1}^{2}=&\sum_{a=1}^{d-1}\left(\prod_{b=1}^{a-1}\sin^{2}
\theta_{b}\right)d\theta_{a}.
\end{align}

A CHS, ${\cal H}^{d}/\Gamma$, is constructed from ${\cal H}^{d}$ 
by acting freely the fundamental group $\Gamma$, a discrete subgroup of 
SO$(d,1)$. In case of $d=2$, a tessellation of the hyperbolic space 
${\cal H}^{2}$ by the tiles of identical regular $2(g+2)$-gon leads to a string
of doughnuts of genus $g$ without boundary.  
Substituting the curvature \eqref{cu1} into the Gauss-Bonnet theorem,
we compute the area $A$ of the obtained orientable CHS, 
\begin{align}
A=4\pi (g-1)\ell^{2},\qquad (g\ge 2).
\label{GB}
\end{align}
For a fixed length scale $\ell$, the large area can be achieved by the large
number of genii $g$.   
When $d\ge 3$~\cite{Thur}, 
CHS's are rigid~\cite{Mosto} that all the metrical quantities 
are determined by the topology and the scale $\ell$, including
their volume, 
\begin{align}
{\rm Vol}({\cal H}^{d}/\Gamma)=e^{\alpha}\ell^{d},
\label{RT}
\end{align}
where $\alpha$ is a topological number fixed by the fundamental group, $\Gamma$,  of the CHS.
Suppose that we have a $d$-dimensional CHS of its maximum distance $2L$. 
Then its volume is smaller than a disk $D$ having a radius $2L$ out of the hyperbolic space,
\begin{align}
{\rm Vol}({\cal H}^{d}/\Gamma)<V(D)\equiv e^{\alpha_{\!{\small D}}}\ell^{d}
=\Omega_d I_d(L/\ell) \ell^d ,
\label{eq:exponent}
\end{align}
where $\Omega_d= \pi^{d/2}/\Gamma(d/2+1)$ is the volume of $d$-sphere and
$I_d(x)=\int_0^x dy \sinh^{d-1}\! y$. For a sufficiently large volume of
$L/\ell\gg 1$,
$\alpha_{\!{\small D}} \approx (d-1)\frac{L}{\ell}+\log \Omega_d /C$ where 
$C=2, 2^3,6, \cdots$ for $d=2,3,4,\cdots$, respectively.  
When $d\ge 3$, the rigidity theorem tells us that $L/\ell$ cannot be chosen 
arbitrarily but determined by the fundamental group of CHS, $\Gamma$.

Though we have two mass scales in this CHS, $1/\ell$ from the constant
curvature and $1/({\rm Vol}_{d})^{\frac{1}{d}}\sim e^{-\frac{\alpha}{d}}/\ell$ 
from the volume, the first eigenmode of Laplace-Beltrami operator is 
mainly decided by $1/\ell$. The bound of the first eigenmode is conjectured to 
be~\cite{Selb} 
\begin{align}
k_{1}\ell\ge 1/2,
\label{eq:lower}
\end{align}
and the number theory approach proved up to 
$k_{1}\ell\ge \sqrt{171/784}\approx 0.22$~\cite{Luo}. (See \cite{Cornish:1997an} for the higher KK excitations.)

If we consider 2-dimensional CHS with a large genus $g$ or 
higher-dimensional $(d\ge 3)$ CHS, we can construct extra dimensions with a large volume but an undetectably large KK mass gap.

For instance, a model with 2-dimensional extra dimensions 
of 1 fm scale thickness and 1 TeV KK mass gap requires $g={\cal O}(10^{6})$ for $d=2$ and $\alpha/d={\cal O}(10)$ for $d\geq 3$.
On the other hand, with $1/\ell \sim 10^{13}$ GeV and $\alpha\sim 23$, we get a sufficiently large volume for a natural scenario for inflation with NM coupling, which we shall show later on.

\section{Inflation with non-minimal coupling}

In this section we briefly recapitulate slow-roll inflation with NM
coupling in $(1+3)$ dimensions and point out the naturalness problem
in fitting the CMB data. In Jordan frame, the action is given as follows:
\begin{align}
S_{{\rm J}}=\int d^4 x\sqrt{-g_{{\rm J}}} \left[-\frac{M^2 + K(\phi)}{2}
R_{{\rm J}}-\tfrac{1}{2} (\partial \phi)^2 -V(\phi)\right],
\label{eq:Jordan}
\end{align}
where $M$ is a mass scale for gravity, $R_{{\rm J}}$ is the Ricci scalar in the Jordan 
frame, $K(\phi)$ is a generic function of $\phi$, and $V(\phi)$ is the 
scalar potential.

One can always find a function $\Omega(\phi)$,
\begin{align}
e^{-2\Omega} = \frac{M_{{\rm Pl}}^2}{M^2+K(\phi)},
\end{align}
so that the Weyl transformation,
\begin{align}
g_{\mu\nu}^{{\rm J}} \to g_{\mu\nu}^{{\rm E}} e^{-2\Omega(\phi)},
\end{align} 
leads to Einstein frame.  
The gravity is canonically normalized as
\begin{align}
S_{{\rm E}} =\int d^4 x\sqrt{-g_{{\rm E}}} 
\left[-\frac{M_{{\rm Pl}}^2 }{2}R-\tfrac{1}{2}(\partial{\hat \phi})^{2}
-U_{{\rm E}}(\hat{\phi}) \right],
\label{Upo}
\end{align}
where the scalar field ${\hat \phi}$ in the Einstein frame is chosen to make
its kinetic term canonical by
\begin{eqnarray}
\frac{d\hat{\phi}}{d\phi}=e^{-\Omega}\sqrt{1 + \frac{3}{2 M_{{\rm Pl}}^2} 
e^{-2\Omega}K'(\phi)^2}\, .
\end{eqnarray}
Subsequently the scalar potential is now read as
\begin{eqnarray}
U_{{\rm E}}(\hat{\phi}) = \frac{M_{{\rm Pl}}^4}{\left[M^2+K(\phi(\hat{\phi}))
\right]^2}\, V(\phi(\hat{\phi})).
\label{Up1}
\end{eqnarray}
When $M\sim M_{{\rm Pl}}$, $U_{{\rm E}}({\hat \phi})\sim V(\phi({\hat \phi}))$
at the small field limit, $\lim_{\phi\rightarrow 0}K\ll M^{2}$.

Suppose $K$ and $V$ satisfy the asymptotic relation, 
$\lim_{\phi\to \infty} (V/K^2) \to C>0$~\cite{history, Park:2008hz}. 
Then the potential \eqref{Up1}
involves a flat plateau at the large field limit, 
$U_{{\rm E}} \stackrel{\phi\rightarrow \infty}{\to} 
M_{{\rm Pl}}^4 V/K^2 \approx M_{{\rm Pl}}^4 C$,
which can be responsible for the slow-roll inflation, and thus 
${\hat \phi}$ can be
identified as an inflaton field~\cite{Park:2008hz}. 

 With $K=\xi_{\rm eff}\phi^2 $ and $V=\lambda_{\rm eff}\phi^4$ ~\cite{Bezrukov:2007ep}, the inflaton potential in Einstein frame  is explicitly read as
\begin{align}
U_{{\rm E}} =\frac{\lambda_{\rm eff}M_4^4}{2 m (\xi_{\rm eff})^2}
\left[1+\frac{M_4^2}{K_4(\phi)}\right]^{-2},
\label{eq:energy}
\end{align}
where $\phi$ is related to the canonically normalized field, ${\hat \phi}$, as
\begin{eqnarray}
\phi \simeq 
\frac{M_4}{\sqrt{\xi_{\rm eff}}}\exp \frac{{\hat \phi}}{\sqrt{6+1/\xi_{\rm eff}} M_4}.
\end{eqnarray}

From the observation of primordial density perturbation by WMAP \cite{wmap7},
\begin{eqnarray}
\delta_H^2 =\frac{1}{75\pi M_4^6}\frac{U_{{\rm E}}^3}{U_{{\rm E}}'}\approx (1.91 \times 10^{-5})^2,
\end{eqnarray}
we get the ratio $\lambda_{\rm eff}/(\xi_{\rm eff})^2$ \cite{Park:2008hz}  
\begin{eqnarray}
\frac{\lambda_{\rm eff}}{(\xi_{\rm eff})^2}\approx 
(2.1 \times 10^{-5})^2. 
\label{eq:effective}
\end{eqnarray}
We regard the appearance of this small number as an unnatural fine-tuning.
In the next section, we address this fine-tuning problem by introducing CHS.

Other cosmological observables, spectral index of scalar perturbation and tensor perturbation, are calculated in terms  of the number of e-foldings and a useful coefficient 
$$a=
1+\frac{1}{6\xi_{\rm eff}}, $$%
as 
\begin{eqnarray}
&&n_{{\rm s}} \approx 1-\frac{2}{N} -\frac{9 a}{2N^2} \approx 0.965\, ,\\
&&r \approx \frac{12 a}{N^2}\approx 0.003,
\end{eqnarray}
which are completely consistent with the current WMAP 7year data \cite{wmap7}.

\section{Inflation with Non-minimal coupling in higher $D$ with compact hyperbolic 
extra dimensions}

Let us consider a model with $d$-dimensional compact hyperbolic extra dimensions.
The model in $D=4+d$ dimensions is described by the action 
\begin{align}
S= \int\!\! d^4 x \hspace{-1mm}\int_{{\cal H}^d/\Gamma} \hspace{-5mm}
d^{d} y \bigg[ -\frac{M_D^{2+d} +K_D}{2}R_D - \tfrac{1}{2}(\partial \phi_D )^2  
- V_D \bigg],
\label{eq:4+d+n}
\end{align}
where $M_D$ is a scale for $D$-dimensional gravity and $K_D$ and 
$V_D$ are (polynormial) functions of $\phi_D$ satisfying $V_D/K_D^2 \to C$ at a large field limit.

If we assume a factorizable geometry,
\begin{eqnarray}
ds_D^2 = ds_4^2 + ds_{{\cal H}^{d}/\Gamma}^2,
\end{eqnarray}
it is straightforward to get the Ricci scalar,
\begin{eqnarray}
R_{D} = R_4 + R_{{\cal H}^{d}/\Gamma}.
\label{Rex}
\end{eqnarray}
As the curvature of the hyperbolic space is negative, the second term in 
\eqref{Rex} can have contribution to the cosmological constant, 
probably negatively, in 4 dimensions. 
Here the cosmological constant in the dimensionally 
reduced effective theory is assumed to be
zero at the minimum of the scalar potential, which is measured 
to be extremely tiny in the present universe. 
In this paper, we will not deal with the
cosmological constant problem but will focus on the slow-roll inflation at the 
large field limit. 
 
Integrating over the extra dimensions of the volume ${\mathbb V}_{d}$ the 
4-dimensional effective action becomes
\begin{eqnarray}
S_4 = \int d^4 x \left[-\frac{M_4^2 + K_4}{2} R_4 - \tfrac{1}{2}(\partial \phi)^{2}
-V_{4}(\phi)\right],
\end{eqnarray}
where $\phi =\sqrt{{\mathbb V}_d}\,\phi_{D}$, 
$M_{\rm Pl}^2\equiv M_4^2 = M_{D}^{2+d} {\mathbb V}_d$, and $K_4 =K_{D} {\mathbb V}_d$. 
This action recovers Eq.~\eqref{eq:Jordan}. Here the possible contribution from KK states could be neglected once the scale from the curvature of CHS is greater than the scale of our interest, the inflation scale, as we have discussed in Eq.~\eqref{eq:lower}.

Without loss of much generality\footnote{The conclusion remains the same as $p=2$ even when we take more general polynomial terms as $K_D\sim \phi_D^p$ and $V_D \sim \phi_D^{2p}$ with $p\in \mathbb{Z}_+$. },  let us consider an explicit case:
$K_D(\phi_D) = \xi_D\phi_D^{2}$ and $V_D(\phi_D)=\lambda_D \phi_D^4$. 
In general, the slow-roll inflation is achieved when $V_D/K_D^2 \to 
\mbox{constant}$, as discussed previously~\cite{Park:2008hz}. 
In $D$ dimensions, $[K_D]=D-2$, $[V_D]=D$, and $[\phi_D]=\tfrac{D-2}{2}$. 
Hence, regardless of $D$, the NM coupling is 
always dimensionless, $[\xi_D ] =0$, but the quartic self-coupling is 
dimensionful except $D=4$ as $[\lambda_D ]=-(D-4)$. 
Here we introduce  
dimensionless couplings $\xi_0\sim 1$ and  $\lambda_0\sim 1$ which count the strength of 
the coupling in the fundamental unit: 
\begin{eqnarray}
\xi_D \equiv \xi_0,\quad \lambda_D  \equiv \frac{\lambda_0}{M_D^{D-4}}.
\end{eqnarray}
Together with $\xi_{{\rm eff}}= \xi$, 
the effective quartic coupling is obtained 
in terms of $\lambda$ as
\begin{eqnarray}
\lambda_{\rm eff} = \frac{\lambda_0}{M_D^{d}{\mathbb V}_d}
=\frac{\lambda_0}{{\cal V}_d}.
\end{eqnarray}
When the scale of $D$-dimensional gravity is the same as the curvature scale
of CHS, $M_{D}~1/\ell$, ${\cal V}_d$ becomes a dimensionless measure of 
the volume of extra dimensions, 
${\cal V}_d=\mbox{Vol}({\cal H}^d/\Gamma)/\ell^{d}$.
An immediate result is
\begin{eqnarray}
\frac{\lambda_{\rm eff}}{(\xi_{\rm eff})^2} \simeq \frac{1}{{\cal V}_d }\times 
\frac{\lambda_0}{\xi_0^2} \ll 1
\end{eqnarray}
provided that the CHS has a large volume, ${\cal V}_d \gg 1$. This is naturally understood as the ${\cal O}(10)$ topological number appears in the exponent which determines the volume of CHS as in Eq. \eqref{eq:exponent}.

For ${\cal V}_d \sim 10^{10}$, the scales in $D$-dimensions are 
$M_{D}\sim 10^{13}$GeV, $\xi_D\sim 1$ and $\lambda_D\sim 10^{-13}$/GeV.
For the mass term of the bulk scalar field, there will be additional 
contribution from the $-K_{D}R_D/2$ term so that we get $m_{{\rm eff}}^2=m^{2}+\frac{d(d-1)\xi_D}{\ell^{2}} \sim 1/\ell^2$ which is of the same scale of the KK mass gap.

\section{Summary and discussion}

A non-minimally coupled scalar field has a potential which has a flat plateau in Einstein frame when the ratio of the square of the NM coupling term and the potential energy is asymptotically constant $V(\phi)/K^2(\phi) \to C$. For this scalar field being an inflaton field fitting CMB data a fine-tuning is required as given in Eq.~\eqref{eq:effective}. In the theory with a single scale, $10^{13}$ GeV,  and NM coupling,  we show that the large volume of CHS, which is controlled by a topological number of the order of a few tens, can provide a plausible resolution of the fine-tuning problem without accompanying a light KK state. In the high scale inflation, reheating takes place at high energy so that the late time cosmology is seamlessly consistent with the current model.  Again, the situation in the current paper is very different from the conventional models with large extra dimensions \cite{ADD} where the low mass gap ($\sim 1/V_n^{1/n}\sim {\rm MeV}$) often spoils the successful slow-roll inflation. The prediction of $r\approx 0.3\%$ level gravitational wave in CMB might be testable in the future experiments such as Planck.

Last but not least some comments on consistency of the current model are in order. 
\begin{itemize}
\item{\it Quantum gravity correction}: One may worry if the energy density of inflaton during the inflation might
be too high  so that it exceeds  the critical density at which black hole 
forms. This is not
the case as is easily checked by Eq.~\eqref{eq:energy}. Indeed, the energy density is highly suppressed by
the large volume %
\begin{eqnarray}
U_{\rm E} \sim \frac{M_4^4}{{\cal V}_d } \ll M_{\rm Pl}^4.
\end{eqnarray}
Thus we can safely neglect quantum gravity effects here.

\item{\it KK correction}: The  de-Sitter temperature during the inflation is certainly less than the scale of the first KK excitation:
\begin{eqnarray}
\frac{H_{{\rm E}}}{2\pi} \approx \sqrt{\frac{U_{{\rm E}}}{12 \pi^2  M_4^2}} \approx \sqrt{\frac{\lambda}{72 \pi^2  m \xi^2}}M_D <\frac{1}{\ell}
\end{eqnarray}
which is well below the first KK scale in  Eq.~\eqref{eq:lower}.
The validity of the purely four dimensional effective description is fully consistent when ${\cal V}_d \gg 1$.

\item{\it Higher order correction}: The current model requires the large field behavior $V/K^2\to C$ to realize the flat potential for slow-rolling. Certainly this might bring unwanted difficulties in understanding effective field theory description since higher order terms $\alpha_n (\phi/M)^n$ with large $n$ may become important at $\phi/M>1$ where inflation took place if the Wilson coefficients of these terms, $\alpha_n$,  are ${\cal O}(1)$. 
This problem is common in large field inflation models such as chaotic inflation model with $~m^2 \phi^2$ or $~\lambda \phi^4$ potential. However, currently we do not have any evidence for $\alpha_n={\cal O}(1)$. For more discussion, see \cite{DeSimone:2008ei}.

\item{\it Astrophysical bounds}: If there remains a light KK graviton like in Ref.~\cite{ADD}, several astrophysical processes including supernova cooling process can be significantly affected. With CHS, on the other hand, the KK scale is high enough not to have a  light graviton mode.  
 
\end{itemize}

\vspace{5mm}

\section*{Acknowledgments}
The authors would like to thank J. Levin, J. McGreevy, 
and D. Orlando for valuable discussions.
Y. Kim is supported by the National Research Foundation of Korea(NRF)
grant funded by the Korea government(MEST) (No. 2009-0062869) through
Astrophysical Research Center for the Structure and Evolution of the Cosmos
(ARCSEC)) and
by the Korea Research Foundation Grant funded by the Korean Government
(KRF-2008-313-C00170).
S. C. Park is supported by the World Premier International Research Center Initiative 
(WPI initiative) by MEXT and also supported by the Grant-in-Aid for scientific 
research (Young Scientists (B) 21740172) from JSPS, Japan.

\end{document}